# A Bayesian Approach to Discovering Truth from Conflicting Sources for Data Integration

Bo Zhao[1]     Benjamin I. P. Rubinstein[2]     Jim Gemmell[2]     Jiawei Han[1]

[1]Department of Computer Science, University of Illinois, Urbana, IL, USA
[2]Microsoft Research, Mountain View, CA, USA

[1]{bozhao3, hanj}@illinois.edu,  [2]{ben.rubinstein, jim.gemmell}@microsoft.com

## ABSTRACT

In practical data integration systems, it is common for the data sources being integrated to provide conflicting information about the same entity. Consequently, a major challenge for data integration is to derive the most complete and accurate integrated records from diverse and sometimes conflicting sources. We term this challenge the *truth finding problem*. We observe that some sources are generally more reliable than others, and therefore a good model of source quality is the key to solving the truth finding problem. In this work, we propose a probabilistic graphical model that can automatically infer true records and source quality without any supervision. In contrast to previous methods, our principled approach leverages a generative process of two types of errors (false positive and false negative) by modeling two different aspects of source quality. In so doing, ours is also the first approach designed to merge multi-valued attribute types. Our method is scalable, due to an efficient sampling-based inference algorithm that needs very few iterations in practice and enjoys linear time complexity, with an even faster incremental variant. Experiments on two real world datasets show that our new method outperforms existing state-of-the-art approaches to the truth finding problem.

## 1. INTRODUCTION

A classic example of data integration is the consolidation of customer databases after the merger of two companies. Today, the scale of data integration has expanded as businesses share all kinds of data in partnership using the Internet, and even more so when information is harvested by search engines crawling billions of Web sites. As data is integrated, it is common for the data sources to claim conflicting information about the same entities. For example, one online book seller may have the complete list of authors for a book, another may only have the first author, while yet another has the wrong author. Consequently, a key challenge of data integration is to derive the most complete and accurate merged records from diverse and sometimes conflicting sources. We term this challenge the *truth finding problem*.

Perhaps the simplest approach to the truth finding problem is majority voting: only treat claims made by a majority of sources as truth. Unfortunately voting can produce false positives if the majority happens to be unreliable; for instance for an obscure fact. This drawback motivates a threshold which a majority proportion of sources must exceed in order for their collective claim to be used. For example we may only reconcile the majority claim if half or more of sources come to a consensus. By varying the threshold we trade false positives for false negatives.

In practice there is no way to select an optimal voting threshold other than applying supervised methods, which are not feasible for large-scale automated data integration. Moreover, voting is effectively *stateless* in that nothing is learned about the reliability of sources from integrating one set of claims to the next; each source is treated equally even if it proves to be unreliable in the long-term.

A better approach to truth finding is to model source quality. Given knowledge of which sources are trustworthy, we can more effectively reconcile contradictory claims by down-weighing the claims of unreliable sources. Conversely, the set of claims consistent with the overall consensus may yield estimates of source quality. Therefore, it is natural to iteratively determine source quality and infer underlying truth together. Specific mechanisms have been proposed in previous work on truth finding [4,7,10,11,14,15], leveraging this principle along with additional heuristics.

While existing methods determine the single most confident truth for each entity, in practice multiple values can be true simultaneously. For example, many books do not have a single author, but instead have a multi-valued author attribute type. Previous approaches are not designed for such real-world settings.

As we shall see, a related drawback of current approaches is that their models of source quality as a single parameter are insufficient, as they overlook the important distinction between two sides of quality. Some sources tend to omit true values, *e.g.*, only representing first authors of a book, individually suffering false negatives; and others introduce erroneous data, *e.g.*, associating incorrect authors with a book, suffering false positives. If for each entity there is only one true fact and each source only makes one claim, then false positives and false negatives are equivalent. However, where multiple facts can be true and sources can make multiple claims per entity, the two types of errors do not necessarily correlate. Modeling these two aspects of source quality separately is the key to naturally allowing multiple truths for each entity, and is a major distinction of this paper.

EXAMPLE 1. *Table 1 shows a sample integrated movie database with movie titles, cast members, and sources. All of the records are correct, except that BadSource.com claims Johnny Depp was in the Harry Potter movie. This false claim can be filtered by majority voting, but then Rupert Grint in the Harry Potter movie will be erroneously treated as false as well. We might try lowering the threshold from 1/2 to 1/3 based on evaluation on costly*

Table 1: An example raw database of movies.

| Entity (Movie) | Attribute (Cast) | Source |
|---|---|---|
| Harry Potter | Daniel Radcliffe | IMDB |
| Harry Potter | Emma Waston | IMDB |
| Harry Potter | Rupert Grint | IMDB |
| Harry Potter | Daniel Radcliffe | Netflix |
| Harry Potter | Daniel Radcliffe | BadSource.com |
| Harry Potter | Emma Waston | BadSource.com |
| Harry Potter | Johnny Depp | BadSource.com |
| Pirates 4 | Johnny Depp | Hulu.com |
| ... | ... | ... |

*labeled data in order to recognize Rupert, but then Johnny will also be treated as true as a consequence. If we knew that Netflix tends to omit true cast data but never includes wrong data, and BadSource.com makes more false claims than IMDB, we may accurately determine the truth. That is, two-sided source quality is needed to make the correct inferences.*

To automatically infer the truth and two-sided source quality, we propose a Bayesian probabilistic graphical model we call the *Latent Truth Model (LTM)* which leverages a generative error process. By treating the truth as a latent random variable, our method can naturally model the complete spectrum of errors and source quality in a principled way—an advantage over heuristics utilized in previous methods. Experiments on two real world datasets—author data from online book sellers and directors from movie sources used in the Bing movies vertical—demonstrate the effectiveness of LTM. We also propose an efficient inference algorithm based on collapsed Gibbs sampling which in practice converges very quickly and requires only linear time with regard to the size of the data.

Our Bayesian model has two additional features. LTM provides a principled avenue for incorporating prior knowledge about data sources into the truth-finding process, which is useful in practice particularly in low data volume settings. Second, if data arrives online as a stream, LTM can learn source quality and infer truth incrementally so that quality learned at the current stage can be utilized for inference on future data. This feature can be used to avoid batch re-training on the cumulative data at each step.

To summarize, our main contributions are as follows:

1. To the best of our knowledge, we are the first to propose a principled probabilistic approach to discovering the truth and source quality simultaneously without any supervision;

2. We are the first to model two-sided source quality, which makes our method naturally support multiple truths for the same entity and achieve more effective truth-finding;

3. We develop an efficient and scalable linear complexity inference algorithm for our model;

4. Our model can naturally incorporate prior domain knowledge of the data sources for low data volume settings; and

5. Our model can run in either batch or online streaming modes for incremental data integration.

In the following sections, we first describe our data model and formalize the problem in Section 2. We then introduce two-sided source quality, the latent truth model and the inference algorithms in Sections 3, 4 and 5. Section 6 presents our experimental results. Several possible improvements of the method and related work are discussed in Sections 7 and 8. Finally, we conclude the paper in Section 9.

## 2. PROBLEM FORMULATION

In general, a data source provides information about a number of attribute types. The quality of a source may be different for each attribute type, for example, an online book seller may be very reliable about authors but quite unreliable about publishers. Thus, each attribute type may be dealt with individually, and for the remainder of this paper we assume a single attribute type is under consideration to simplify the discussion.[1]

We now provide the details of our data model, and formally define the truth finding problem.

### 2.1 Data Model

We assume a single, multi-valued attribute type, for example authors of a book, or cast of a movie. The input data we consume is in the form of triples *(entity, attribute, source)* where *entity* serves as a key identifying the entity, *attribute* is one of possibly many values for the given entity's attribute type, and *source* identifies from where the data originates. This representation can support a broad range of structured data, such as the movie data shown in Table 1. For the sake of examining claims made about attributes by different sources, we re-cast this underlying input data into tables of facts (distinct attribute values for a given entity) and claims (whether each source did or did not assert each fact) as follows.

*Definition 1.* Let $\mathcal{DB} = \{row_1, row_2, ..., row_N\}$ be the *raw database* we take as input, where each row is in the format of $(e, a, c)$, where $e$ is the entity, $a$ is the attribute value, and $c$ is the source. Each row is unique in the raw database.

Table 1 is an example of a raw database.

*Definition 2.* Let $\mathcal{F} = \{f_1, f_2, ..., f_F\}$ be the set of distinct *facts* selected from the raw database, each fact $f$ is an entity-attribute pair with an id as the fact's primary key: $(id_f, e_f, a_f)$. The entity-attribute pair in each row of the fact table should be unique (while the pair may appear in multiple rows of the raw database).

Table 2 is an example of the fact table obtained from the raw database in Table 1.

Table 2: The fact table of Table 1.

| FID | Entity (Movie) | Attribute (Cast) |
|---|---|---|
| 1 | Harry Potter | Daniel Radcliffe |
| 2 | Harry Potter | Emma Waston |
| 3 | Harry Potter | Rupert Grint |
| 4 | Harry Potter | Jonny Depp |
| 5 | Pirates 4 | Jonny Depp |
| ... | ... | ... |

*Definition 3.* Let $\mathcal{C} = \{c_1, c_2, ..., c_C\}$ be the set of *claims* generated from the raw database. Each claim $c$ is in the format of $(f_c, s_c, o_c)$, where $f_c$ is the id of the fact associated with the claim, $s_c$ is the source of the claim, and $o_c$ is the observation of the claim, taking a Boolean value True or False.

Specifically, for each fact $f$ in the fact table:

1. For each source $s$ that is associated with $f$ in the raw database, we generate a *positive* claim: $(id_f, s, True)$, meaning source $s$ asserted fact $f$.

---

[1]LTM can integrate each attribute type in turn, and can be extended to use global quality (*cf.* Section 7).

2. For each source $s$ that is not associated with $f$, but is associated with the entity in fact $f$, i.e., $e_f$, in the raw database, we generate a *negative* claim: $(id_f, s, False)$, meaning source $s$ did not assert fact $f$ but asserted some other facts associated with entity $e_f$.

3. For other sources that are not associated with entity $e_f$ in the raw database, we do not generate claims, meaning those sources do not have claims to make with regard to entity $e_f$.

Moreover, we denote the set of claims that are associated with fact $f$ as $\mathcal{C}_f$, and the rest of the claims as $\mathcal{C}_{-f}$. And let $\mathcal{S} = \{s_1, s_2, ..., s_S\}$ be the set of sources that appear in $\mathcal{C}$, let $\mathcal{S}_f$ be the set of sources that are associated with fact $f$ in $\mathcal{C}$, and $\mathcal{S}_{-f}$ be the set of sources not associated with $f$.

Table 3 is an example of the claim table generated from the raw database in Table 1. IMDB, Netflix and BadSource.com all asserted that Daniel was an actor in the Harry Potter movie, so there is a positive claim from each source for fact 1. Netflix did not assert Emma was an actress of the movie, but since Netflix did assert other cast members of the movie, we generate a negative claim from Netflix for fact 2. Since Hulu.com did not assert any cast members of the Harry Potter movie, we do not generate claims from Hulu.com for any facts associated with the movie.

**Table 3: The claim table of Table 1.**

| RID | Source | Observation |
|---|---|---|
| 1 | IMDB | True |
| 1 | Netflix | True |
| 1 | BadSource.com | True |
| 2 | IMDB | True |
| 2 | Netflix | False |
| 2 | BadSource.com | True |
| 3 | IMDB | True |
| 3 | Netflix | False |
| 3 | BadSource.com | False |
| 4 | IMDB | False |
| 4 | Netflix | False |
| 4 | BadSource.com | True |
| 5 | Hulu.com | True |
| ... | ... | ... |

*Definition 4.* Let $\mathcal{T} = \{t_1, t_2, ..., t_T\}$ be a set of *truths*, where each truth $t$ is a Boolean value taking True/False and is associated with one fact in $\mathcal{F}$, indicating whether this fact is true or not. For each $f \in \mathcal{F}$, we denote the truth associated with $f$ as $t_f$.

Note that we are not given the truths in the input. Instead, we must infer the hidden truths by fitting a model and generating the truth table. For the sake of evaluation here, a human generated truth table is compared to algorithmically generated truth tables.

Table 4 is a possible truth table for the raw database in Table 1. Daniel Radcliffe, Emma Watson and Rupert Gint are the actual cast members of Harry Potter, while Jonny Depp is not, and therefore the observations for facts 1,2,3 are True, and False for fact 4.

### 2.2 Problem Definitions

We can now define the problems of interest in this paper.

**Inferring fact truth.** Given an input raw database $\mathcal{DB}$ with no truth information, we want to output the inferred truths $\mathcal{T}$ for all facts $\mathcal{F}$ contained in $\mathcal{DB}$.

**Table 4: A truth table for raw database Table 1.**

| RID | Entity (Movie) | Attribute (Cast) | Truth |
|---|---|---|---|
| 1 | Harry Potter | Daniel Radcliffe | True |
| 2 | Harry Potter | Emma Waston | True |
| 3 | Harry Potter | Rupert Grint | True |
| 4 | Harry Potter | Jonny Depp | False |
| 5 | Pirates 4 | Jonny Depp | True |
| ... | ... | ... | ... |

**Inferring source quality.** Besides the truth of facts, we also want to automatically infer *quality* information for each source represented in $\mathcal{DB}$. Source quality information indicates how reliable each source is for the given attribute type. Source quality information can be used for understanding data sources, selecting good sources in order to produce more accurate truth, uncovering or diagnosing problems with crawlers, providing prior knowledge for inferring truth from new data, etc.

Fact truth and source quality are not independent; they are closely related, and, in fact, are computed simultaneously by our principled approach. The quality of a source is used to help decide whether to believe its given claims, and the correctness of a source's claims can be used to determine the source's quality.

We formally introduce our measures of source quality in the next section. Subsequently, we will explain how learning the quality of sources and truth of facts is naturally integrated in LTM.

## 3. TWO-SIDED SOURCE QUALITY

In this section, we examine how to measure source quality in our truth discovery model and why quality measures utilized in previous work are inadequate in practice.

### 3.1 Revisiting Quality Measures

We can treat each source as a classifier on facts in the sense that each source makes true or false claims/predictions for the facts. Thus given ground truth for a subset of facts, we can grade the quality of the sources by looking at how close their predictions are to the ground truth. Similarly, our measures apply to truth finding mechanisms which we treat as ensembles of source classifiers.

Based on the observation of claims and truth of facts for each source $s$, we produce the source's confusion matrix in Table 5 ($o$ stands for observation and $t$ stands for truth) and several derivative quality measures as follows.

**Table 5: Confusion matrix of source $s$.**

|  | $t = True$ | $t = False$ |
|---|---|---|
| $o = True$ | True Positives ($TP_s$) | False Positives ($FP_s$) |
| $o = False$ | False Negatives ($FN_s$) | True Negatives ($TN_s$) |

- *Precision* of source $s$ is the probability of its positive claims being correct, i.e., $\frac{TP_s}{TP_s + FP_s}$.

- *Accuracy* of source $s$ is the probability of its claims being correct, i.e., $\frac{TP_s + TN_s}{TP_s + FP_s + TN_s + FN_s}$.

- *Sensitivity* or *Recall* of source $s$ is the probability of true facts being claimed as true, i.e., $\frac{TP_s}{TP_s + FN_s}$. And $1 - sensitivity$ is known as the *false negative rate*.

- *Specificity* of source $s$ is the probability of false facts being claimed as false, i.e., $\frac{TN_s}{FP_s + TN_s}$. And $1 - specificity$ is known as the *false positive rate*.

Table 6 presents the different source quality measures computed for the three movie sources from the example Claim Table (Table 3) and example Truth Table (Table 4).

Table 6: Quality of sources based on Tables 3 and 4.

| Measures | IMDB | Netflix | BadSource.com |
|---|---|---|---|
| TP | 3 | 1 | 2 |
| FP | 0 | 0 | 1 |
| FN | 0 | 2 | 1 |
| TN | 1 | 1 | 0 |
| Precision | 1 | 1 | 2/3 |
| Accuracy | 1 | 1/2 | 1/2 |
| Sensitivity | 1 | 1/3 | 2/3 |
| Specificity | 1 | 1 | 0 |

## 3.2 Limitations of Precision

Some previous works [10, 11, 14] use the single metric of precision for modeling the quality of sources, which means they only consider positive claims while ignoring negative claims. Those methods should not have trouble deciding the *single* most confident true fact, but when multiple facts can be true for each entity, and some of them have less support, measuring source quality by precision alone cannot utilize the value of negative claims to recognize erroneous data. The following example illustrates this limitation:

EXAMPLE 2. *In Table 6, BadSource.com has 2 true positives out of three positive claims, for a fair precision of 2/3. As a result, BadSource.com's false claim (Harry Potter, Johnny Depp) may still be given some credence and even be regarded as true, taking advantage of the true positive claims made by BadSource.com. However, if we consider negative claims, and know the quality of negative claims, we could mitigate the erroneous inference. For example, if we know IMDB has perfect recall and its negative claims are always correct, we can easily detect that (Harry Potter, Johnny Depp) is not true since IMDB claims it is false.*

## 3.3 Limitations of Accuracy

In order to avoid the problems posed by only considering positive claims, recent work [7] has taken negative claims into consideration. However, the adopted approach still only measures the quality of sources as a single value: accuracy.

Any single value for quality overlooks two fundamentally different types of errors: false positives and false negatives, which are not necessarily correlated. For example it is possible that a source has very high precision but very low recall, resulting in a fairly low accuracy. The low accuracy would let us discount the source omitting a value; but we would also be forced to discount a positive the source claims, even though it has perfect precision. Put another way, *a scalar-valued measure forces us to treat a low precision source exactly like a low recall source.*

EXAMPLE 3. *In Table 6 we can see that Netflix makes more false negatives than BadSource.com, but makes no false positives. However, by making one more true positive claim than Netflix, BadSource.com can gain exactly the same accuracy as Netflix. In this situation, the true positive claims made by Netflix will be affected by its high false negative rate, while, on the other hand, the low false negative rate of BadSource.com could lead to false information being introduced. By using only accuracy to judge the quality of sources while inferring truth, a positive claim by BadSource.com will be treated just as trustworthy as one from Netflix, despite the difference in their precision. So, if we want to be able to accept attributes about an entity that only Netflix knows about (which seems reasonable, given its perfect precision), we would be forced to accept attributes about an entity known only to BadSource.com (which is risky, given its low precision).*

## 3.4 Sensitivity and Specificity

Clearly, the use of a scalar-valued quality value can never capture the two error types, false positives and false negatives, which have different implications for data integration. A very conservative source would only make claims it is very certain of, yielding few false positives but many false negatives. On the other hand, a venturous source may have very few false negatives while frequently making erroneous claims.

Therefore, in contrast with all the previous methods, we model the *sensitivity* and *specificity* of sources as two independent quality measures. With sensitivity associated with false negatives and specificity associated with false positives, we are able to cover the complete spectrum of source quality. In the next section we will explain how we model the two quality signals as two independent random variables that have different prior probabilities in order to simulate real-world scenarios. The following example illustrates the advantages of modeling source quality with these two metrics:

EXAMPLE 4. *In Table 6 sensitivity and specificity reveal more details of the error distribution of each source. From these, we see that Netflix has low sensitivity and high specificity, so we will give less penalty to the facts (Harry Potter, Emma Watson) and (Harry Potter, Rupert Grint) that are claimed false by it, while still giving (Harry Potter, Daniel Radcliffe), which it claims as true, higher confidence. Additionally, we know that BadSource.com has low specificity and IMDB has high sensitivity, so we see that (Harry Potter, Johny Depp) is likely false given that it is claimed true by BadSource.com and claimed false by IMDB.*

The only question that remains is how to model the sensitivity and specificity of sources without knowing the truth of facts. So far, our examples have implicitly assumed a supervised setting where ground truth is known and is used to calculate quality measures, while in practice unsupervised methods that do not have such knowledge are required. In the next section, we will introduce our proposed approach, the Latent Truth Model (LTM), which naturally solves the problem by treating both truth and quality as latent random variables, so that in each iteration of inference, we will first have the truth information available so that we can calculate the source quality based on it, then we go back and re-infer truth based on updated source quality. By introducing the latent truth, our method can model the relation between truth and source quality in a principled way, rather than utilizing heuristics as in previous methods.

## 4. LATENT TRUTH MODEL

In this section we will formally introduce our proposed model, called the Latent Truth Model, for discovering the truth of facts and the quality of data sources. We will first give a brief introduction to Bayesian networks, then discuss the intuitions behind our model, briefly explain the major components of the approach and how it can model our intuitions, and finally we provide details about how the model is constructed.

### 4.1 Review of Bayesian Networks

The Bayesian Network is a powerful formalism for modeling real-world events based on prior belief and knowledge of conditional independence [12]. A Bayesian network is a directed acyclic

553

probabilistic graphical model in the Bayesian sense: each node represents a random variable, which could be observed values, latent (unobserved) values, or unknown parameters. A directed edge from node $a$ to $b$ ($a$ is then called the parent of $b$) models the conditional dependence between $a$ and $b$ in the sense that the random variable associated with a child node follows a probabilistic conditional distribution that takes values depending on the parent nodes as parameters.

Given the observed data and prior and conditional distributions, various inference algorithms can perform *maximum a posteriori (MAP)* estimation to assign latent variables and unknown parameters values that (approximately) maximize the posterior likelihoods of those corresponding unobserved variables given the data.

Bayesian networks have been proven to be effective in numerous tasks such as information extraction, clustering, text mining, etc. In this work, our proposed Latent Truth Model is a new Bayesian network for inferring truth and source quality for data integration.

## 4.2 Intuition Behind the Latent Truth Model

We next describe the intuition behind modeling the quality of sources, truth of facts and claim observations as random variables, before detailing the LTM graphical model in the next section.

### 4.2.1 Quality of Sources

As discussed in the previous section, we need to model the quality of sources as two independent factors: specificity and sensitivity, and therefore in our model we create two separate random variables for each source, one associated with its specificity and the other with its sensitivity.

Moreover, in practice we often have prior belief or assumptions with regard to the data sources. For example, it is reasonable to assume that the majority of data coming from each source is not erroneous, *i.e.*, the specificity of data sources should be reasonably high. On the other hand, we could also assume missing data is fairly common, *i.e.*, sensitivity may not be high for every source. It is also possible that we have certain prior knowledge about the quality of some *specific* data sources that we want to incorporate into the model. In all these cases, the model should be able to allow us to plug in such prior belief. For this reason, in LTM we model source quality in the Bayesian tradition so that any available assumptions or domain knowledge can be easily incorporated by specifying prior distributions for the source quality variables. In the absence of such knowledge, we can simply use uniform priors.

### 4.2.2 Truth of Facts

In LTM we model the probability of (or belief in) each fact being true as an unknown or latent random variable in the unit interval. In addition, we also introduce the actual truth label of each fact, which depends on the represented probability, as a latent Boolean random variable. By doing so, at any stage of the computation we can clearly distinguish the two types of errors (false positives and false negatives) so that the specificity and sensitivity of sources can be modeled in a natural and principled way.

In addition, if we have any prior belief about how likely all or certain specific facts are true, our model can also support this information by setting prior distributions for the truth probability. Otherwise, we use a uniform prior.

### 4.2.3 Observation of Claims

Now we need to model our actual observed data: claims from different sources. Recall that each claim has three components: the fact it refers to, the source it comes from and the observation (True/False). Clearly, the observation of the claim depends on two factors: whether the referred fact is indeed true or false, and what is the quality of the data source asserting the claim. In particular, if the fact is false, then a high specificity of the source indicates the observation is more likely to also be false, while a low specificity (or high false positive rate) means the observation is more likely to be true; on the other hand, if the fact is true, then a high sensitivity of the source implies the observation is more likely to be true, and otherwise low sensitivity means the claim is more likely to be a false negative.

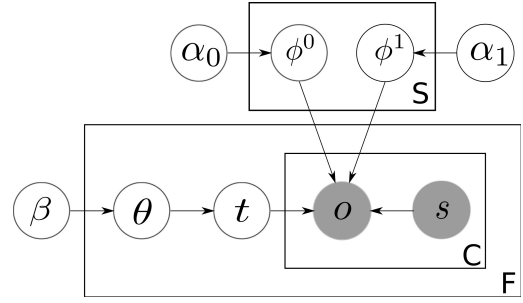

**Figure 1: The probabilistic graphical model of LTM.**

As we can see, with latent truth and two-sided source quality, all four possible real-world outcomes can be simulated naturally. We must model the observations of claims as random variables which depend on the truth of their referred facts and the quality of their sources. Then given the actual claim data, we can go back and infer the most probable fact truth and source quality (effectively inverting the directions of edges via Bayes rule). And by controlling the observation altogether, the latent truth and the source quality can mutually influence each other through the *joint* inference, in the sense that claims produced by high quality sources are more likely to be correct and sources that produce more correct claims are more likely to be high quality.

## 4.3 Model Details

Now we will explain the details of LTM. Figure 1 shows the graphical structure of conditional dependence of our model. Each node in the graph represents a random variable or prior parameter, and darker shaded nodes indicate the corresponding variable is observed (and lighter nodes represent latent variables). A plate with a set as its label means that the nodes within are replicated for each element in the set, *e.g.*, the $\mathcal{S}$ plate indicates that each source has conditionally independent quality nodes.

A directed edge from $a$ to $b$ means $b$ is generated from a distribution that takes values of $a$ as parameters in addition to parameters from the other parents of $b$. The detailed generative process is as follows.

**1. FPR.** For each source $k \in \mathcal{S}$, generate its false positive rate $\phi_k^0$, which is exactly $(1 - specificity)$, from a Beta distribution with hyperparameter $\boldsymbol{\alpha_0} = (\alpha_{0,1}, \alpha_{0,0})$, where $\alpha_{0,1}$ is the prior false positive count, and $\alpha_{0,0}$ is the prior true negative count of each source:

$$\phi_k^0 \sim Beta(\alpha_{0,1}, \alpha_{0,0}).$$

Note that here we model the false positive rate only to make it easier to explain the model in the future, but there is no difference to modeling specificity directly.

The Beta distribution is utilized because it is the conjugate prior of Bernoulli and Binomial distributions—those distributions used below for children nodes—and inference is more efficient as a result. Its parameter $\boldsymbol{\alpha_0}$ controls the prior belief for sensitivity of

554

sources, and in practice, we set $\alpha_{0,0}$ significantly higher than $\alpha_{1,0}$ to plug in our assumptions that sources in general are good and do not have high false positive rate, which is not only reasonable but also important since otherwise the model could flip every truth while still achieving high likelihood thereby making incorrect inferences.

**2. Sensitivity.** For each source $k \in \mathcal{S}$, generate its sensitivity $\phi_k^1$ from a Beta distribution with hyperparameter $\boldsymbol{\alpha_1} = (\alpha_{1,1}, \alpha_{1,0})$, where $\alpha_{1,1}$ is the prior true positive count, and $\alpha_{1,0}$ is the false negative count of each source:

$$\phi_k^1 \sim Beta(\alpha_{1,1}, \alpha_{1,0}) \,.$$

Similar to $\boldsymbol{\alpha_0}$ above, $\boldsymbol{\alpha_1}$ controls the prior distribution for sensitivity of sources. Since in practice we observe that it is quite common for some sources to ignore true facts and therefore generate false negative claims, we will not specify a strong prior for $\boldsymbol{\alpha_1}$ as we do for $\boldsymbol{\alpha_0}$, instead we can just use a uniform prior.

**3. Per fact.** For each fact $f \in \mathcal{F}$,

**3(a). Prior truth probability.** Generate prior truth probability $\theta_f$ from a Beta distribution with hyperparameter $\boldsymbol{\beta} = (\beta_1, \beta_0)$, where $\beta_1$ is the prior true count, and $\beta_0$ is the prior false count of each fact:

$$\theta_f \sim Beta(\beta_1, \beta_0) \,.$$

Here $\boldsymbol{\beta}$ determines the prior distribution of how likely each fact is to be true. In practice, if we do not have a strong belief, we can use a uniform prior meaning it is equally likely to be true or false and the model can still effectively infer the truth from other factors in the model.

**3(b). Truth label.** Generate the truth label $t_f$ from a Bernoulli distribution with parameter $\theta_f$:

$$t_f \sim Bernoulli(\theta_f) \,.$$

As a result, $t_f$ is a Boolean variable, and the prior probability that $t_f$ is true is exactly $\theta_f$.

**3(c). Observation.** For each claim $c$ of fact $f$, i.e., $c \in \mathcal{C}_f$, denote its source as $s_c$, which is an observed dummy index variable that we use to select the corresponding source quality. We generate the observation of $c$ from a Bernoulli distribution with parameter $\phi_{s_c}^{t_f}$, i.e., quality parameter of source $s_c$ depending on $t_f$, the truth of $f$:

$$o_c \sim Bernoulli(\phi_{s_c}^{t_f}) \,.$$

Specifically, if $t_f = 0$, then $o_c$ is generated from a Bernoulli distribution with parameter $\phi_{s_c}^0$, i.e., the false positive rate of $s_c$, as:

$$o_c \sim Bernoulli(\phi_{s_c}^0) \,.$$

Then the resulting value of $o_c$ is Boolean. If it is true then the claim is a false positive claim and its probability is exactly the false positive rate of $s_c$.

If $t_f = 1$, $o_c$ is generated from a Bernoulli distribution with parameter $\phi_{s_c}^1$, i.e., the sensitivity of $s_c$:

$$o_c \sim Bernoulli(\phi_{s_c}^1) \,.$$

Then in this case the probability that the Boolean variable $o_c$ is true is exactly the sensitivity or true positive rate of $s_c$ as desired.

## 5. INFERENCE ALGORITHMS

In this section we discuss how to perform inference to estimate the truth of facts and quality of sources from the model, given the observed claim data.

### 5.1 Likelihood Functions

According to the Latent Truth Model, the probability of each claim $c$ of fact $f$ given the LTM parameters is:

$$p(o_c|\theta_f, \phi_{s_c}^0, \phi_{s_c}^1) = p(o_c|\phi_{s_c}^0)(1 - \theta_f) + p(o_c|\phi_{s_c}^1)\theta_f \,.$$

Then the complete likelihood of all observations, latent variables and unknown parameters given the hyperparameters $\boldsymbol{\alpha_0}, \boldsymbol{\alpha_1}, \boldsymbol{\beta}$ is:

$$p(\boldsymbol{o},\boldsymbol{s},\boldsymbol{t},\boldsymbol{\theta},\boldsymbol{\phi^0},\boldsymbol{\phi^1}|\boldsymbol{\alpha_0},\boldsymbol{\alpha_1},\boldsymbol{\beta}) = \prod_{s \in \mathcal{S}} p(\phi_s^0|\boldsymbol{\alpha_0})p(\phi_s^1|\boldsymbol{\alpha_1}) \times$$

$$\times \prod_{f \in \mathcal{F}} \left( p(\theta_f|\boldsymbol{\beta}) \sum_{t_f \in 0,1} \theta_f^{t_f}(1-\theta_f)^{1-t_f} \prod_{c \in \mathcal{C}_f} p(o_c|\phi_{s_c}^{t_f}) \right) \,. \quad (1)$$

### 5.2 Truth via Collapsed Gibbs Sampling

Given observed claim data, we must find assignments of latent truth that maximize the joint probability, i.e., get the *maximum a posterior* (MAP) estimate for $\boldsymbol{t}$:

$$\hat{\boldsymbol{t}}_{MAP} = \arg\max_{\boldsymbol{t}} \int \int \int p(\boldsymbol{o},\boldsymbol{s},\boldsymbol{t},\boldsymbol{\theta},\boldsymbol{\phi^0},\boldsymbol{\phi^1}) \mathrm{d}\boldsymbol{\theta}\mathrm{d}\boldsymbol{\phi^0}\mathrm{d}\boldsymbol{\phi^1} \,.$$

As we can see, a brute force inference method that searches the space of all possible truth assignment $\boldsymbol{t}$ would be prohibitively inefficient. So we need to develop a much faster inference algorithm.

Gibbs sampling is a Markov chain Monte Carlo (MCMC) algorithm that can estimate joint distributions that are not easy to directly sample from. The MCMC process is to iteratively sample each variable from its conditional distribution given all the other variables, so that the sequence of samples forms a Markov chain, the stationary distribution of which is just the exact joint distribution we want to estimate.

Moreover, LTM utilizes the conjugacy of exponential families when modeling the truth probability $\boldsymbol{\theta}$, source specificity $\boldsymbol{\phi^0}$ and sensitivity $\boldsymbol{\phi^1}$, so that they can be integrated out in the sampling process, i.e., we can just iteratively sample the truth of facts and avoid sampling these other quantities, which yields even greater efficiency. Such a sampler is commonly referred to as a *collapsed* Gibbs sampler.

Let $\boldsymbol{t}_{-f}$ be the truth of all facts in $\mathcal{F}$ except $f$. We iteratively sample for each fact given the current truth labels of other facts:

$$p(t_f = i|\boldsymbol{t}_{-f}, \boldsymbol{o}, \boldsymbol{s}) \propto \beta_i \prod_{c \in \mathbf{C}_f} \frac{n_{s_c,i,o_c}^{-f} + \alpha_{i,o_c}}{n_{s_c,i,1}^{-f} + n_{s_c,i,0}^{-f} + \alpha_{i,1} + \alpha_{i,0}} \quad (2)$$

where

$$n_{s_c,i,j}^{-f} = |\{c' \in \mathcal{C}_{-f}|s_{c'} = s_c, t_{f_{c'}} = i, o_{c'} = j\}| \,,$$

i.e., the number of $s_c$'s claims whose observation is $j$, and referred fact is not $f$ and its truth is $i$. These counts reflect the quality of $s_c$ based on claims of facts other than $f$, e.g., $n_{s_c,0,0}^{-f}$ is the number of true negative claims of $s_c$, $n_{s_c,0,1}^{-f}$ is the false positive count, $n_{s_c,1,0}^{-f}$ is the false negative count, and $n_{s_c,1,1}^{-f}$ is the true positive count.

The detailed derivation of Equation (2) can be found in Appendix A. Intuitively, it can be interpreted as sampling the truth of each fact based on the prior for truth and the quality signals of associated sources estimated on other facts.

Algorithm 1 presents pseudo-code for implementing the collapsed Gibbs sampling algorithm. We initialize by randomly assigning each fact a truth value, and calculate the initial counts for each source. Then in each iteration, we re-sample each truth variable

**Algorithm 1** Collapsed Gibbs Sampling for Truth

{Initialization}
**for all** $f \in \mathcal{F}$ **do**
  {Sample $t_f$ from uniform}
  **if** $random() < 0.5$ **then**
    $t_f \leftarrow 0$
  **else**
    $t_f \leftarrow 1$
  **for all** $c \in \mathcal{C}_f$ **do**
    $n_{s_c, t_f, o_c} \leftarrow n_{s_c, t_f, o_c} + 1$
{Sampling}
**for** $i = 1 \rightarrow K$ **do**
  $i \leftarrow i + 1$
  **for all** $f \in \mathcal{F}$ **do**
    $p_{t_f} \leftarrow \beta_{t_f}, p_{1-t_f} \leftarrow \beta_{1-t_f}$
    **for all** $c \in \mathcal{C}_f$ **do**
      $p_{t_f} \leftarrow \frac{p_{t_f} \times (n_{s_c, t_f, o_c} - 1 + \alpha_{t_f, o_c})}{n_{s_c, t_f, 1} + n_{s_c, t_f, 0} - 1 + \alpha_{t_f, 1} + \alpha_{t_f, 0}}$
      $p_{1-t_f} \leftarrow \frac{p_{1-t_f} \times (n_{s_c, 1-t_f, o_c} + \alpha_{1-t_f, o_c})}{n_{s_c, 1-t_f, 1} + n_{s_c, 1-t_f, 0} + \alpha_{1-t_f, 1} + \alpha_{1-t_f, 0}}$
    {Sample $t_f$ from conditional distribution}
    **if** $random() < \frac{p_{1-t_f}}{p_{t_f} + p_{1-t_f}}$ **then**
      $t_f \leftarrow 1 - t_f$
      {$t_f$ changed, update counts}
      **for all** $c \in \mathcal{C}_f$ **do**
        $n_{s_c, 1-t_f, o_c} \leftarrow n_{s_c, 1-t_f, o_c} - 1$
        $n_{s_c, t_f, o_c} \leftarrow n_{s_c, t_f, o_c} + 1$
    {Calculate expectation of $t_f$}
    **if** $i > burnin$ and $i\%thin = 0$ **then**
      $p(t_f = 1) \leftarrow p(t_f = 1) + t_f / samplesize$

from its distribution conditioned on all the other truth variables, and the quality counts for each source will be updated accordingly.

For final prediction, we could use samples in the last round, or a more stable method is to calculate the expectation of truth for each fact in the way that we discard the first $m$ samples (*burn-in period*) then for every $n$ sample in the remainder we calculate their average (*thinning*), which is to prevent correlation in the samples. Then if the expectation is equal to or above a threshold of 0.5, we predict the fact is true, otherwise it is false.

The time complexity of Algorithm 1 is $O(|\mathcal{C}|)$ or $O(|\mathcal{S}| \times |\mathcal{F}|)$, which is linear in the number of claims. Comparing to a brute force search algorithm with complexity $O(2^{|\mathcal{F}|})$, our collapsed Gibbs sampling method is much more efficient and scalable.

### 5.3 Estimating Source Quality

After we obtain the predictions of fact truth, we can immediately read off the source quality signals from LTM by treating the truth as observed data.

A maximum a posterior (MAP) estimate of source quality has a closed-form solution since the posterior of source quality is also a Beta distribution:

$$sensitivity(s) = \phi_s^1 = \frac{E[n_{s,1,1}] + \alpha_{1,1}}{E[n_{s,1,0}] + E[n_{s,1,1}] + \alpha_{1,0} + \alpha_{1,1}},$$

$$specificity(s) = 1 - \phi_s^0 = \frac{E[n_{s,0,0}] + \alpha_{0,0}}{E[n_{s,0,0}] + E[n_{s,0,1}] + \alpha_{0,0} + \alpha_{0,1}}$$

where $E[n_{s,i,j}] = \sum_{c \in \mathcal{C}, s_c=s, o_c=j} p(t_{f_c} = i)$ is the expected quality counts of source $s$ which depends on the truth probability of each fact $s$'s claims output by Algorithm 1. These counts also allow us to estimate other quality measures of sources, *e.g.*, precision:

$$precision(s) = \frac{E[n_{s,1,1}] + \alpha_{1,1}}{E[n_{s,0,1}] + E[n_{s,1,1}] + \alpha_{0,1} + \alpha_{1,1}}.$$

An advantage of MAP estimation on the LTM graphical model is that we can incorporate prior knowledge with regard to specific sources or all data sources together.

### 5.4 Incremental Truth Finding

If input data arrives online as a stream, we can use the source quality learned at the current stage as the prior for future data. Incrementally learning on new data involves essentially the same algorithm as the batch setting. Specifically, for each source we use $E[n_{s,i,j}] + \alpha_{i,j}$ as its quality prior to replace $\alpha_{i,j}$, and fit LTM only on the new data. Thus fitting LTM can be achieved online with complexity at each step as above but only in terms of the size of the increment to the dataset.

A simpler and more efficient approach is to assume that the source quality remains relatively unchanged over the medium term; then we can directly compute the posterior truth probability of each fact as

$$p(t_f = 1 | \boldsymbol{o}, \boldsymbol{s}) = \frac{\beta_1 \prod_{c \in \mathbf{C}_f} (\phi_s^1)^{o_c} (1 - \phi_s^1)^{1-o_c}}{\sum_{i=0,1} \beta_i \prod_{c \in \mathbf{C}_f} (\phi_s^i)^{o_c} (1 - \phi_s^i)^{1-o_c}}. \quad (3)$$

Periodically the model can then be retrained batch-style on the total cumulative data, or incrementally on the data arrived since the model was last updated.

## 6. EXPERIMENTS

In this section, we demonstrate the effectiveness of our method compared with state-of-the-art algorithms on two real world datasets. In addition to assessing statistical performance, we also conduct efficiency experiments that show that our model converges quickly in practice and that our inference algorithm is scalable.

### 6.1 Experimental Setup

#### 6.1.1 Datasets

We use the two following real world datasets and generate one synthetic dataset to stress test the effectiveness of our method when source quality is low.

**Book Author Dataset.** This data, crawled from abebooks.com consists of 1263 book entities, 2420 book-author facts, and 48153 claims from 879 book seller sources. 100 books were randomly sampled and their true authors were manually labeled. While this dataset has been used previously [4, 14], there all the authors observed by each source for the same book were concatenated as one single claim. In this work, we substantially clean the data and segment the authors, since our method can naturally handle multiple truth attributes for the same entity.

**Movie Director Dataset.** This data, used in the Bing movies vertical for surfacing reviews, meta-data and entity actions such as "rent" and "stream", consists of 15073 movie entities, 33526 movie-director facts, and 108873 claims from 12 sources enumerated in Table 8. 100 movies were randomly sampled for their true directors to be manually labeled. Our original dataset contained more movies, but to make this dataset more difficult and interesting, we removed those movies that only have one associated director or only appear in one data source, *i.e.*, we only keep the conflicting records in our database.

Table 7: Inference results per dataset and per method with threshold 0.5.

|  | Results on book data | | | | | Results on movie data | | | | |
| --- | --- | --- | --- | --- | --- | --- | --- | --- | --- | --- |
|  | One-sided error | | | Two-sided error | | One-sided error | | | Two-sided error | |
|  | Precision | Recall | FPR | Accuracy | F1 | Precision | Recall | FPR | Accuracy | F1 |
| **LTMinc** | **1.000** | 0.995 | **0.000** | **0.995** | **0.997** | 0.943 | 0.914 | 0.150 | **0.897** | **0.928** |
| **LTM** | **1.000** | 0.995 | **0.000** | **0.995** | **0.997** | 0.943 | 0.908 | 0.150 | 0.892 | 0.925 |
| 3-Estimates | **1.000** | 0.863 | **0.000** | 0.880 | 0.927 | 0.945 | 0.847 | 0.133 | 0.852 | 0.893 |
| Voting | **1.000** | 0.863 | **0.000** | 0.880 | 0.927 | 0.855 | 0.908 | 0.417 | 0.821 | 0.881 |
| TruthFinder | 0.880 | **1.000** | 1.000 | 0.880 | 0.936 | 0.731 | **1.000** | 1.000 | 0.731 | 0.845 |
| Investment | 0.880 | **1.000** | 1.000 | 0.880 | 0.936 | 0.731 | **1.000** | 1.000 | 0.731 | 0.845 |
| LTMpos | 0.880 | **1.000** | 1.000 | 0.880 | 0.936 | 0.731 | **1.000** | 1.000 | 0.731 | 0.845 |
| HubAuthority | **1.000** | 0.322 | **0.000** | 0.404 | 0.488 | **1.000** | 0.620 | **0.000** | 0.722 | 0.765 |
| AvgLog | **1.000** | 0.169 | **0.000** | 0.270 | 0.290 | **1.000** | 0.025 | **0.000** | 0.287 | 0.048 |
| PooledInvestment | **1.000** | 0.142 | **0.000** | 0.245 | 0.249 | **1.000** | 0.025 | **0.000** | 0.287 | 0.048 |

**Synthetic Dataset.** We follow the generative process described in Section 4 to generate this synthetic dataset. There are 10000 facts, 20 sources, and for simplicity each source makes a claim with regard to each fact, *i.e.*, 200000 claims in total. To test the impact of sensitivity, we set expected specificity to be 0.9 ($\alpha_0 = (10, 90)$), and vary expected sensitivity from 0.1 to 0.9 ($\alpha_0$ from $(10, 90)$ to $(90, 10)$), and use each parameter setting to generate a dataset. We do the same for testing the impact of specificity by setting $\alpha_1 = (90, 10)$ and varying $\alpha_0$ from $(90, 10)$ to $(10, 90)$. In all datasets $\beta = (10, 10)$.

### 6.1.2 Environment

All the experiments presented were conducted on a workstation with 12GB RAM, Intel Xeon 2.53GHz CPU, and Windows 7 Enterprise SP1 installed. All the algorithms including previous methods were implemented in C# 4.0 and complied by Visual Studio 2010.

## 6.2 Effectiveness

We compare the effectiveness of our latent truth model (LTM) and the incremental version LTMinc and a truncated version LTMpos with several previous methods together with voting. We briefly summarize them as follows, and refer the reader to the original publications for details.

LTMinc. For each dataset, we run standard LTM model on all the data except the 100 books or movies with labeled truth, then apply the output source quality to predict truth on the labeled data using Equation (3) and evaluate the effectiveness.

LTMpos. To demonstrate the value of negative claims, we run LTM only on positive claims and call this truncated approach LTMpos.

Voting. For each fact, compute the proportion of corresponding claims that are positive.

TruthFinder [14]. Consider positive claims only, and for each fact calculate the probability that at least one positive claim is correct using the precision of sources.

HubAuthority, AvgLog [10] [11]. Perform random walks on the bipartite graph between sources and facts constructed using only positive claims. The original HubAuthority (HITS) algorithm was developed to compute quality for webpages [9], AvgLog is a variation.

Investment, PooledInvestment [10] [11]. At a high level, each source uniformly distributes its credits to the attributes it claims positive, and gains credits back from the confidence of those attributes.

3-Estimates [7]. Negative claims are considered, and accuracy is used to measure source quality. The difficulty of data records is also considered when calculating source quality.

Parameters for the above algorithms are set according to the optimal settings suggested by their authors. For our method, as we previously explained, we need to set a reasonably high prior for specificity, *e.g.*, 0.99, and the actual prior counts should be at the same scale as the number of facts to become effective, which means we set $\alpha_0 = (10, 1000)$ for book data, and $(100, 10000)$ for movie data. For other prior parameters, we just use a small uniform prior, which means we do not enforce any prior bias. Specifically we set $\alpha_1 = (50, 50)$ and $\beta = (10, 10)$ for both datasets.

### 6.2.1 Quantitative Evaluation of Truth Finding

All algorithms under comparison can output a probability for each fact indicating how likely it is to be true. Without any supervised training, the only reasonable threshold probability is 0.5. Table 7 compares the effectiveness of different methods on both datasets using a 0.5 threshold.

As we can see, both the accuracy and F1 score of LTM (and LTMinc) are significantly better than the other approaches on both datasets. On the book data we almost achieve perfect performance. The performance on the movie data is lower than the book data because we intentionally make the movie data more difficult. There is no significant difference between the performance of LTM and LTMinc, which shows that source quality output by LTM is effective for making incremental truth prediction on our datasets. For simplicity we will only mention LTM in the comparison of effectiveness with other methods in the remainder of this section.

Overall 3-Estimates is the next best method, demonstrating the advantage of considering negative claims. However, since that approach uses accuracy to measure source quality, some negative claims could be trusted more than they should be. Therefore, although it can achieve high precision, even greater than our method on the movie data, this algorithm's recall is fairly low, resulting in worse overall performance than LTM.

Voting achieves reasonably good performance on both datasets as well. Its precision is perfect on books but its recall is lower, since that dataset on average has more claims on each fact and therefore attributes that have majority votes are very likely to be true. However, many sources only output first authors, so the other authors cannot gather enough votes and will be treated as false. On the more difficult movie data, Voting achieves higher recall than precision, this is because there are fewer sources in this dataset and therefore false attributes can more easily gain half or more votes. In this case it is necessary to model source quality.

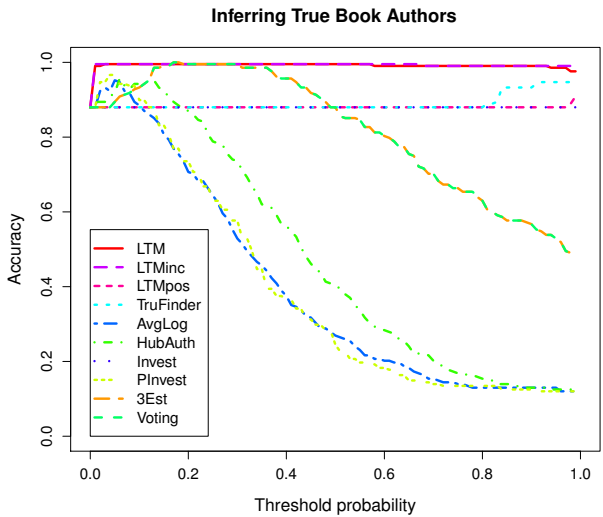
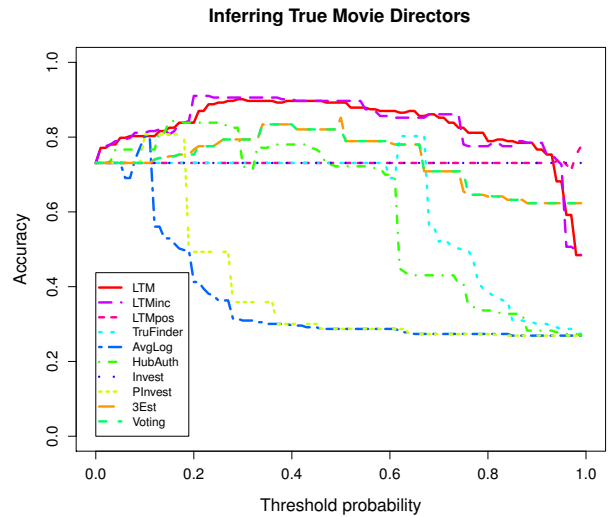

Figure 2: Accuracy vs. thresholds on the book data and the movie data.

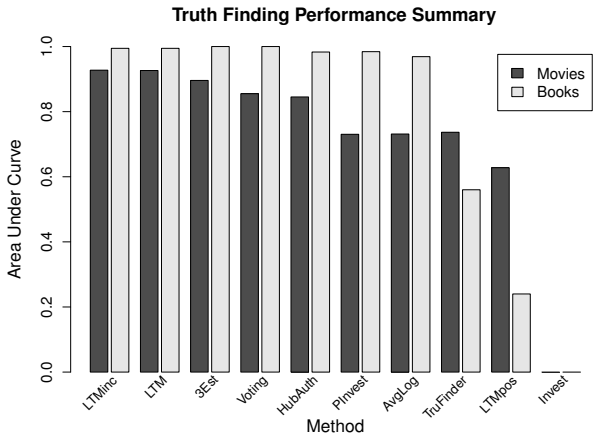

Figure 3: AUCs per method per dataset, sorted by decreasing average AUC.

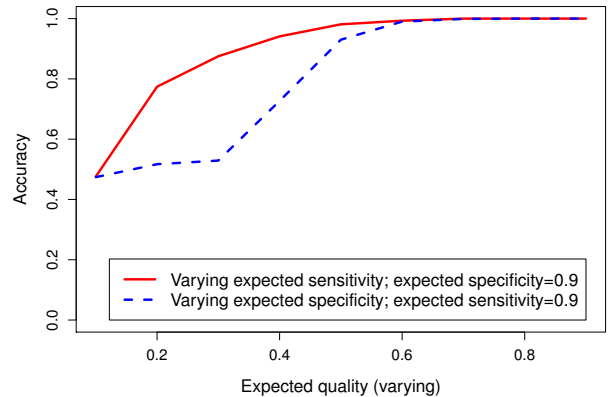

Figure 4: **LTM** under varying expected synthetic source quality (sensitivity/specificity; with the other's expectation fixed).

Note that in our experiments Voting achieves better performance than it appears to achieve in previous work. In previous experiments, votes are calculated on concatenated attribute lists rather than individual attributes. For example, if the author list $\langle a, b \rangle$ gets 2 votes, and the list $\langle a, c \rangle$ gets 3 votes, then author $a$ should actually get 5 votes. In previous settings, comparisons with Voting are not completely fair.

TruthFinder, Investment and LTMpos appear too optimistic in their prediction, since their 1.0 false positive rate on both datasets implies they are predicting everything to be true. This is expected since TruthFinder uses the probability that at least one positive claim is correct to predict truth, which may work to find the most likely truth but will not be sufficiently discriminative if multiple attributes can be simultaneously true. Without considering negative claims, LTMpos also fails as expected, which further proves it is critical to consider negative claims when multiple truths are possible.

On the other hand, HubAuthority, AvgLog and PooledInvestment all seem to be overly conservative. They all have perfect precision but their recall is fairly low on both datasets, resulting in overall lowest accuracy and F1.

Next we demonstrate how the algorithms' performances change as we vary the threshold probability for truth. This illuminates the distributions of probability scores assigned by each algorithm. Note that although in practice there is no good way to select the optimal threshold other than performing supervised training, it is still of interest to view each method's performance at their optimal threshold if training data were available. A more confident algorithm would assign true records higher probability and false records lower probability, so that the performance would be more stable with regard to the threshold.

The first sub-figure of Figure 2 plots the accuracy versus threshold on the book data; the plot of F1 is omitted since it looks very similar with an almost identical shape to each curve. We can see that LTM is quite stable no matter where the threshold is set, indi-cating our method can discriminate between true and false better than other methods. Voting and 3-Estimates are rather conservative, since their optimal threshold is around 0.2, where their performance is even on par with our method. However, in practice it is difficult to find such an optimal threshold. Their performance drops very fast when the threshold increases above 0.5, since more false negatives are produced. The optimal threshold for HubAuthority, AvgLog, and PooledInvestment are even lower and their performance drops even faster when the threshold increases, indicating they are more conservative by assigning data lower probability than deserved. On the other hand, TruthFinder, Investment and LTMpos are overly optimistic. We can see the optimal threshold for TruthFinder is around 0.95, meaning its output scores are too high. Investment and LTMpos consistently think everything is true even at a higher threshold.

The second sub-figure of Figure 2 is the analogous plot on the movie data, which is more difficult than the book data. Although LTM is not as stable as on the book data, we can see that it is still consistently better than all the other methods in the range from 0.2 to 0.9, clearly indicating our method is more discriminative and stable. 3-Estimates achieves its optimal threshold around 0.5, and Voting has its peak performance around 0.4, which is still worse than LTM, indicating source quality becomes more important when conflicting records are more common. For other methods, PooledInvestment and AvgLog are still rather conservative, while Investment and LTMpos continue to be overly optimistic. However, it seems TruthFinder and HubAuthority enjoy improvements on the movie data.

Next in Figure 3 we show the area under the ROC curve (AUC) metric of each algorithm on both datasets, which summarizes the performance of each algorithm in ROC space and quantitatively evaluates capability of correctly ranking random facts by score. We can see several methods can achieve AUC close to the ideal of 1 on the book data, indicating that the book data would be fairly easy given training data. On the movie data, however, LTM shows clear advantage over 3-Estimates, Voting and the other methods. Overall on both datasets our method is the superior one.

Last but not least, we would like to understand LTM's behavior when source quality degrades. Figure 4 shows the accuracy of LTM on the synthetic data when the expected specificity or sensitivity of all sources is fixed while the other measure varies between 0.1 and 0.9. We can see the accuracy stays close to 1 until the source quality starts to drop below 0.6, and it decreases much faster with regard to specificity than sensitivity. This shows LTM is more tolerant of low sensitivity, which proves to be effective in practice and is an expected behavior since the chosen priors incorporate our belief that specificity of sources is usually high but sensitivity is not. When specificity is around 0.3 (respectively sensitivity is around 0.1), the accuracy drops to around 0.5 which means the prediction is nearly random.

### 6.2.2 Case Study of Source Quality Prediction

Having evaluated the performance of our model on truth finding, we may now explore whether the source quality predicted by our method is reasonable, bearing in mind that no ground truth is available with which to quantitatively validate quality. Indeed this exercise should serve as a concrete example of what to expect when reading off source quality (*cf.* Section 5.3).

Table 8 shows a MAP estimate of the sensitivity and specificity of sources from our model fit to the movie data, sorted by sensitivity. This table verifies some of our observations on the movie sources: IMDB tends to output rather complete records, while LTM assigns IMDB correspondingly high sensitivity. Note that we can also observe in this table that sensitivity and specificity do not necessarily correlate. Some sources can do well or poorly on both metrics, and it is also common for more conservative sources to achieve lower sensitivity but higher specificity (Fandango), while more aggressive sources to get higher sensitivity but lower specificity (IMDB). This further justifies the intuition that we ought to model the quality of sources as two independent factors.

Table 8: Source quality on the movie data.

| Source | Sensitivity | Specificity |
|---|---|---|
| imdb | **0.911622836** | 0.898838631 |
| netflix | 0.894019034 | 0.934833904 |
| movietickets | 0.862889367 | 0.978844687 |
| commonsense | 0.809752315 | 0.982347827 |
| cinemasource | 0.794184357 | 0.985847745 |
| amg | 0.776583683 | 0.690600694 |
| yahoomovie | 0.760589896 | 0.897654374 |
| msnmovie | 0.749192861 | 0.987870636 |
| zune | 0.744272491 | 0.973922421 |
| metacritic | 0.678661638 | 0.987957893 |
| flixster | 0.584223615 | 0.911078627 |
| fandango | 0.499623726 | **0.989836274** |

### 6.3 Efficiency

We now study the scalability of LTM and LTMinc.

### 6.3.1 Convergence Rate

Since our inference algorithm is an iterative method, we now explore how many iterations it requires in practice to reach reasonable accuracy. To evaluate convergence rate, in the same run of the algorithm, we make 7 sequential predictions using the samples in the first 7, 10, 20, 50, 100, 200, 500 iterations, with burn in iterations 2, 2, 5, 10, 20, 50, 100, and sample gap 0, 0, 0, 1, 4, 4, 9 respectively. We repeat 10 times to account for randomization due to sampling, and calculate the average accuracy and 95% confidence intervals on the 10 runs for each of the 7 predictions, as shown in Figure 5. One can see that accuracy quickly reaches 0.85 even after only 7 iterations, although in the first few iterations mean accuracy increases and variation decreases, implying that the algorithm has yet to converge. After only 50 iterations, the algorithm achieves optimal accuracy and extremely low variation, with additional iterations not improving performance further. Thus we conclude that LTM inference converges quickly in practice.

### 6.3.2 Runtime

We now compare the running time of LTM and LTMinc with previous methods. Although it is easy to see that our algorithms and previous methods all have linear complexity in the number of claims in the data, we expected from the outset that our more principled approach LTM would take more time since it is more complex and requires costly procedures such as generating a random number for each fact in each iteration. However, we can clearly see its effective, incremental version LTMinc is much more efficient without needing any iteration. In particular we recommend that in efficiency-critical situations, standard LTM be infrequently run offline to update source quality and LTMinc be deployed for online prediction.

We created 4 smaller datasets by randomly sampling 3k, 6k, 9k, and 12k movies from the entire 15k movie dataset and by pulling all facts and claims associated with the sampled movies. We then ran each algorithm 10 times on the 5 datasets, for which the average

559

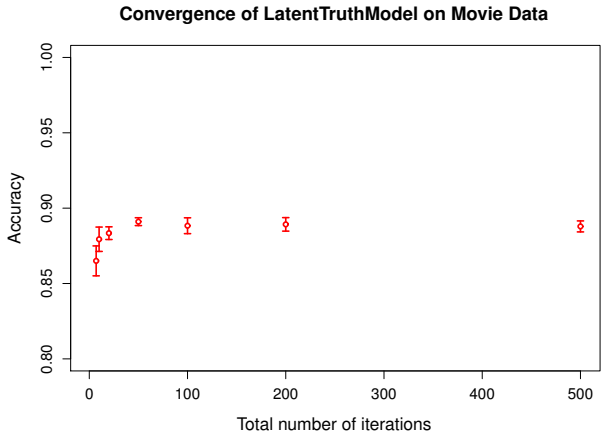

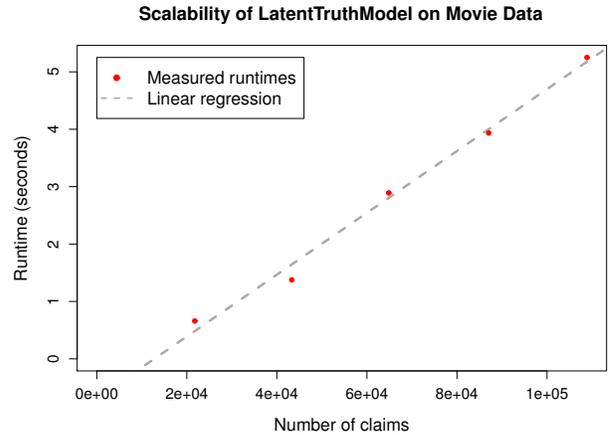

Figure 5: Convergence of **LTM** with 10 repeats per point: error bars show the sample mean and 95% confidence intervals. Accuracy (for threshold 0.5) is truncated at 0.8.

Figure 6: Measurements of the runtime for 100 iterations of **LTM** for varying numbers of claims. The included linear regression enjoys an exceptional goodness-of-fit of $R^2 = 0.9913$.

running times are recorded in Table 9. Note that all the algorithms are iterative except Voting and LTMinc, so for fairness we conservatively fix their number of iterations to 100; and we run LTMinc on the same data as other algorithms by assuming the data is incremental and source quality is given. As we can see, complex algorithms like 3-Estimates and LTM take longer, but only by a factor of 3–5 times the other algorithms. Given the superior accuracy of LTM, we believe the additional computation will usually be acceptable. Moreover, we can see LTMinc is much more efficient than most methods and is almost as efficient as Voting.

Table 9: Comparing runtimes on the movie data.

|  | *Runtimes (secs.) vs. #Entities* | | | | |
|---|---|---|---|---|---|
| *#Entities* | 3k | 6k | 9k | 12k | 15k |
| **Voting** | **0.004** | **0.008** | **0.012** | **0.027** | **0.030** |
| LTMinc | **0.004** | **0.008** | **0.012** | 0.037 | 0.048 |
| **AvgLog** | 0.150 | **0.297** | 0.446 | **0.605** | **0.742** |
| **HubAuthority** | 0.149 | **0.297** | **0.445** | 0.606 | 0.743 |
| PooledInvestment | 0.175 | 0.348 | 0.514 | 0.732 | 0.856 |
| TruthFinder | 0.195 | 0.393 | 0.587 | 0.785 | 0.971 |
| Investment | 0.231 | 0.464 | 0.690 | 0.929 | 1.143 |
| 3-Estimates | 0.421 | 0.796 | 1.170 | 1.579 | 1.958 |
| LTM | 0.660 | 1.377 | 2.891 | 3.934 | 5.251 |

To further verify LTM runs linearly in the number of claims, we perform linear regression on the running time as a function of dataset size (*cf.* Figure 6), which yields an exceptional goodness-of-fit $R^2$ score of 0.9913. This establishes the scalability of LTM.

## 7. DISCUSSIONS

We now revisit the assumptions made by LTM and list several directions for extension to more general scenarios.

**Multiple attribute types.** We have assumed that quality of a source across different attribute types is independent and therefore can be inferred individually. We can, however, extend LTM to handle multiple attribute types in a joint fashion. For each source we can introduce source-specific quality priors $\alpha_{0,s}$ and $\alpha_{1,s}$, which can be regularized by a global prior, and use the same prior to generate type-specific quality signals. Then at each step we can also optimize the likelihood with regard to $\alpha_{0,s}$ and $\alpha_{1,s}$ using Newton's method, so that quality of one attribute type will affect the inferred quality of another via their common prior.

**Entity-specific quality.** LTM assumes a constant quality over all entities presented by a source, which may not be true in practice. For example, IMDB may be accurate with horror movies but not dramas. In response, we can further add an entity-clustering layer to the multi-typed version of LTM discussed above by introducing cluster labels for entities and generate quality signals for each cluster. We can then jointly infer the best partition of entities and cluster-specific quality.

**Real-valued loss.** LTM's loss is either 0 (no error) or 1 (error), but in practice loss can be real-valued, *e.g.*, inexact matches of terms, numerical attributes, *etc*. In such cases a principled truth-finding model similar to LTM, could use *e.g.*, a Gaussian to generate observations from facts and source quality instead of the Bernoulli.

**Adversarial sources.** LTM assumes that data sources have reasonable specificity and precision, *i.e.*, there are few adversarial sources whose majority data are false. However, in practice such sources may exist and their malicious data will artificially increase the specificity of benign sources, causing false data presented by benign sources to become more difficult to detect. Since false facts provided by adversarial sources can be successfully recognized by LTM due to their low support, we can address this problem by iteratively running LTM, at each step removing (adversarial) sources with inferred specificity and precision below some threshold.

## 8. RELATED WORK

There are several previous studies related to the truth finding problem. Resolving inconsistency [1] and modeling source quality [6] have been discussed in the context of data integration. Later [14] was the first to formally introduce the truth-finding problem and propose an iterative mechanism to jointly infer truth and source quality. Then [10] developed several new algorithms and applied integer programming to enforce constraints on truth data, *e.g.*, asserting that city populations should increase with time; [11] designed a framework that can incorporate background information such as how confidently records are extracted from sources; and [7] observed that the difficulty of merging data records should be considered in modeling source quality, in the sense that sources would not gain too much credit from records that are fairly easy to integrate. [13] proposed an EM algorithm for truth finding in sensor networks, but the nature of claims and sensor quality in their setting is rather different than here. In this paper we implement most of the previous algorithms except those models for handling information not available in our datasets, *e.g.*, constraints on truths; and we show that our proposed method outperforms the previous approaches.

Past work also focuses on other aspects, or different data types, in data integration. The copying relationship between sources was studied in [3–5]. By detecting the copying relationship, the support for erroneous data can be discounted and accuracy for truth finding can be improved. [3] also showed it is beneficial to consider multiple attributes together rather than independently. [15] explored semi-supervised truth finding by utilizing the similarity between data records. [2] modeled source quality as relevance to desired queries in a deep web source selection setting. [8] focused on finding truth from several knowledge bases.

## 9. CONCLUSIONS

In this paper, we propose a probabilistic graphical model called the Latent Truth Model to solve the truth finding problem in data integration. We observe that in practice there are two types of errors, false positive and false negative, which do not necessarily correlate, especially when multiple facts can be true for the same entities. By introducing the truth as a latent variable, our Bayesian approach can model the generative error process and two-sided source quality in a principled fashion, and can naturally support multiple truths as a result. Experiments on two real world datasets demonstrate the clear advantage of our method over the state-of-the-art truth finding methods. A case-study of source quality predicted by our model also verifies our intuition that two aspects of source quality should be considered. An efficient inference algorithm based on collapsed Gibbs sampling is developed, which is shown through experiments to converge quickly and cost linear time with regard to data size. Additionally, our method can naturally incorporate various prior knowledge about the distribution of truth or quality of sources, and it can be employed in an online streaming setting for incremental truth finding, which we prove to be much more efficient and as effective as batch inference. We also list several future directions to improve LTM for handling more general scenarios.

## 10. ACKNOWLEDGEMENTS

We thank Ashok Chandra, Duo Zhang, Sahand Negahban and three anonymous reviewers for their valuable comments. The work was supported in part by the U.S. Army Research Laboratory under Cooperative Agreement No. W911NF-09-2-0053 (NS-CTA).

## APPENDIX
## A. DETAILS OF INFERENCE

We can apply Bayes rule to rewrite the conditional distribution of $t_f$ given $\boldsymbol{t}_{-f}$ and the observed data as follows:

$$p(t_f = i | \boldsymbol{t}_{-f}, \boldsymbol{o}, \boldsymbol{s})$$
$$\propto p(t_f = i | \boldsymbol{t}_{-f}) \prod_{c \in \mathbf{C}_f} p(o_c, s_c | t_f = i, \boldsymbol{o}_{-f}, \boldsymbol{s}_{-f}) \ . \quad (4)$$

We first rewrite the first term in Equation (4):

$$p(t_f = i | \boldsymbol{t}_{-f}) = \int p(t_f = i | \theta_f) p(\theta_f | \boldsymbol{t}_{-f}) \mathrm{d}\theta_f$$
$$= \frac{1}{\mathbf{B}(\beta_1, \beta_0)} \int \theta_f^{\beta_1 + i - 1} (1 - \theta_f)^{\beta_0 + (1-i) - 1} \mathrm{d}\theta_f$$
$$= \frac{\mathbf{B}(\beta_1 + i, \beta_0 + (1-i))}{\mathbf{B}(\beta_1, \beta_0)} = \frac{\beta_i}{\beta_1 + \beta_0} \propto \beta_i \ .$$

For the remaining terms in Equation (4), we have:

$$p(o_c, s_c | t_f = i, \boldsymbol{o}_{-f}, \boldsymbol{s}_{-f})$$
$$\propto \int p(o_c | \phi^i_{s_c}) p(\phi^i_{s_c} | \boldsymbol{o}_{-f}, \boldsymbol{s}_{-f}) \mathrm{d}\phi^i_{s_c}$$
$$\propto \int p(o_c | \phi^i_{s_c}) p(\phi^i_{s_c}) \prod_{c' \notin \mathbf{C}_f, s_{c'} = s_c} p(o_{c'} | \phi^i_{s_c}) \mathrm{d}\phi^i_{s_c}$$
$$\propto \frac{\int (\phi^i_{s_c})^{o_c + n^{-f}_{s_c, i, 1} + \alpha_{i,1} - 1} (1 - \phi^i_{s_c})^{(1 - o_c) + n^{-f}_{s_c, i, 0} + \alpha_{i,0} - 1} \mathrm{d}\phi^i_{s_c}}{\mathbf{B}(n^{-f}_{s_c, i, 1} + \alpha_{i,1}, n^{-f}_{s_c, i, 0} + \alpha_{i,0})}$$
$$= \frac{n^{-f}_{s_c, i, o_c} + \alpha_{i, o_c}}{n^{-f}_{s_c, i, 1} + n^{-f}_{s_c, i, 0} + \alpha_{i,1} + \alpha_{i,0}} \ .$$

Now we can incorporate the above two equations into Equation (4) to yield Equation (2).